\documentclass[english]{article}
\usepackage{mathpazo}
\usepackage[T1]{fontenc}
\usepackage[utf8]{inputenc}
\usepackage{geometry}
\usepackage{babel}
\usepackage{textcomp}
\usepackage{amssymb}
\usepackage{graphicx}
\usepackage[authoryear]{natbib}
\usepackage{subscript}
\usepackage[unicode=true,pdfusetitle,
 bookmarks=true,bookmarksnumbered=false,bookmarksopen=false,
 breaklinks=true,pdfborder={0 0 0},pdfborderstyle={},backref=false,colorlinks=false]
 {hyperref}

\makeatletter

\providecommand{\tabularnewline}{\\}

\newcommand{\lyxaddress}[1]{
	\par {\raggedright #1
	\vspace{1.4em}
	\noindent\par}
}

\makeatother

\begin{document}
\title{Advanced testing of low, medium and high ECS CMIP6 GCM simulations
versus ERA5-T2m}
\date{March 25, 2022}
\author{Nicola Scafetta\textsuperscript{1}{*}}
\maketitle

\lyxaddress{\textsuperscript{1}Department of Earth Sciences, Environment and
Georesources, University of Naples Federico II, Complesso Universitario
di Monte S. Angelo, via Cinthia, 21, 80126 Naples, Italy.}

\lyxaddress{{*} Corresponding author: nicola.scafetta@unina.it}

\subsection*{Keypoints}
\begin{itemize}
\item The CMIP6 GCMs are very different from each other; their ECS large
range needs to be greatly constrained for scientific and policy purposes.
\item We group the GCMs into 3 classes with low, medium and high ECS; we
test them against the ERA5-T2m record using spatial t-statistics.
\item We found that the high and medium-ECS GCMs run too hot; the low-ECS
GCMs are not optimal yet, but also unalarming for the future.
\end{itemize}

\subsection*{Citation}

Scafetta, N. (2022). Advanced testing of low, medium, and high ECS
CMIP6 GCM simulations versus ERA5-T2m. \emph{Geophysical Research
Letters}, \textbf{49}, e2022GL097716. \href{https://doi.org/10.1029/2022GL097716}{https://doi.org/10.1029/2022GL097716}

\newpage{}
\begin{abstract}
The equilibrium climate sensitivity (ECS) of the CMIP6 global circulation
models (GCMs) varies from 1.83 °C to 5.67 °C. Herein, 38 GCMs are
grouped into three ECS classes (low, 1.80-3.00 °C; medium, 3.01-4.50
°C; high, 4.51-6.00 °C) and compared against the ERA5-T2m records
from 1980-1990 to 2011-2021. We found that all models with ECS > 3.0
°C overestimate the observed global surface warming and that spatial
t-statistics rejects the data-model agreement over 60\% (using low-ECS
GCMs) to 81\% (using high-ECS GCMs) of the Earth's surface. Thus,
the high and medium-ECS GCMs are unfit for prediction purposes. The
low-ECS GCMs are not fully satisfactory yet, but they are found unalarming
because by 2050 they predict a moderate warming ($\Delta T_{preindustrial\rightarrow2050}\lesssim2\:{^\circ}C$).
\end{abstract}

\subsection*{Plain language summary}

The last-generation CMIP6 GCMs are used by scientists and policymakers
to interpret past and future climatic changes and to determine appropriate
(adaptation or mitigation) policies to optimally address scenario-related
climate-change hazards. However, these models are affected by large
uncertainties. For example, their ECS varies from 1.83 to 5.67 °C,
which makes their 21st-century predicted warming levels very uncertain.
This issue is here addressed by testing the GCMs' global and local
performance in predicting the 1980-2021 warming rates against the
ERA5-T2m record and by grouping them into three ECS classes (low-ECS,
1.80-3.00 °C; medium-ECS, 3.01-4.50 °C; high-ECS, 4.51-6.00 °C). We
found that: (1) all models with ECS > 3.0 °C overestimate the observed
global surface warming; (2) Student t-tests show model failure over
60\% (low-ECS) to 81\% (high-ECS) of the Earth's surface. Thus, the
high and medium-ECS GCMs do not appear to be consistent with the observations
and should not be used for implementing policies based on their scenario
forecasts. The low-ECS GCMs perform better, although not optimally;
however, they are also found unalarming because for the next decades
they predict moderate warming: $\Delta T_{preindustrial\rightarrow2050}\lesssim2\:{^\circ}C$.

\section{Introduction}

The World Climate Research Programme (WCRP) Coupled Model Intercomparison
Projects (CMIP) global climate models (GCMs) are used for interpreting
and forecasting climate change \citep{IPCC2007,IPCC-2,IPCC2021}.
The latest  CMIP6 GCMs are still very different from each other since
their equilibrium climate sensitivity (ECS) -- the equilibrium warming
induced by doubling the atmospheric CO\textsubscript{2} concentration
from 280 to 560 ppm -- ranges between 1.83 to 5.67 °C \citep{IPCC2021}.
Consequently, they predict that the global surface temperature could
warm between 1.0 and 3.3 °C above the pre-industrial period (1850-1900)
even if the anthropocentric emissions stopped today \citep{Huntingford}.

The large ECS uncertainty is due to the poor physical understanding
of various feedback mechanisms such as water vapor and cloudiness
\citep{Knutti,M=0000F6ller1963,IPCC2021}. Some studies suggested
that high-ECS values are not supported by the observations \citep{Nijsse,Jimenez,Zelinka,Zhu};
other studies supported only low ECS values, e.g. 0.5-2.5 °C \citep{Bates,Christy2017,Happer,Kluft,Lewis,Lindzen,McKitrick(2020),Monckton,Smirnov,Stefani}.
Indeed, decadal and millennial climatic oscillations \citep{Alley,Christiansen,Esper,Kutschera,Ljungqvist,Matskovsky,Moberg,Scafetta2014,Scafettasss}
and additional solar/astronomical forcings are still debated \citep{Connolly,Scafetta2004,Scafetta2006,Scafetta2012,Scafetta2013,Scafetta2021,Scafettagg,Svensmark}.
Most GCMs overestimate the surface warming observed since 1980 \citep{Scafetta2013,Scafetta2021,ScafettaMDPI2021,Tokarska,Wang}
as well as that observed in the global \citep{McKitrick(2020)} and
tropical troposphere \citep{Mitchell}, in particular, at its top
(200-300 hPa) where they predict an unobserved hot-spot \citep{McKitrick}.
The above uncertainties prevent an accurate evaluation of the ECS.

Given the large ECS range and subsequent predictions of the GCMs,
having information as to which models may be given more credibility
would be of enormous value for policy. Furthermore, identifying the
regions where the models most disagree from the data is important
for improving them or correcting the data.

\citet{Scafetta2021} tested several CMIP6 GCMs against some temperature
records and found that the data-model agreement improves for the models
with lower ECS. Herein, we provide a complementary and more robust
statistical approach by grouping the same models into three sub-ensembles
according to their ECS: low-ECS, 1.8-3.0 °C; medium-ECS, 3.01-4.50
°C; and high-ECS, 4.51-6.0 °C. We also adopt spatial Student t-statistics,
which can optimally empathize regional dynamical divergences between
observations and a specific ensemble of model predictions. Thus, the
discrepancies being discussed cannot be interpreted simply as model
noise, but should represent a significant model category failure covering
the Earth’s surface or specific regions over the historical period.
Furthermore, we briefly discuss the implication of the findings regarding
the reliability of the 21st-century predictions according to various
emission scenarios.

\section{Data and method}

The monthly reanalysis fields ERA5 Near-Surface Air Temperature (T2m)
record \citep{Huang} from 1980 to June 2021 and the surface air temperature
(tas) records from 1850 to 2100 by 38 different CMIP6 GCMs (Table
\ref{tab1}) were downloaded from the KNMI Climate Explorer \citep{Oldenborgh}.

The ERA5-T2m record was preferred to other options (e.g., HadCRUT,
GISTEM and NOAA surface temperature records) because it covers the
entire world surface and can be more properly used to test the simulations
over the  globe. The condition of spatial completeness is necessary
because the global surface average temperature  records will be also
tested; this operation requires that both observed and modeled records
are obtained by integrating over the same areas. The alternative temperature
records poorly cover the polar, great forest, and desert regions for
the lack of instrumental data \citep[cf.][]{ScafettaMDPI2021}. Their
adoption would require specific model integrations over the same areas
covered by the data but this exercise is left to a future work.

The satellite lower tropospheric temperature (TLT) measurements by
Microwave Sounding Units (MSUs) \citep{Spencer,Mears} were not considered
because their trends should be slightly scaled down to simulate the
surface temperature  \citep{Christy2018,McKitrick(2020),Mitchell}.
However, land-use and urban-changes impact the observed T2m records
and, consequently, the actual global and local climatic warming trends
may be significantly lower than those given by ERA5-T2m  \citep[cf.:][]{Connolly,Tole,Ouyang2019,Scafetta2021.2}.

Simulations using  historical forcings (1850-2014) and four shared
socioeconomic pathway (SSP) scenarios (2015-2100) are available: SSP126
(low GHG emissions), SSP245 (intermediate emissions), SSP370 (high
emissions), and SSP585 (very high emissions). The SSP curves are compatible
with each other up to 2021. GCM tas levels from January 2011 to June
2021 were calculated by averaging the periods 2011-2020 and 2011-2021.

For each grid cell $j=1,...,M$, and for an ensemble of models $i=1,...,N$,
we obtained the observed ($\Delta T_{j}^{o}$) and modeled ($\Delta T_{i,j}^{m}$)
temperature changes by comparing the 2011-2021  to the 1980-1990 means.
Within each grid cell, the differences of the means between the models
and observations are computed as: 

\begin{equation}
\Delta T_{j}=\frac{1}{N}\sum_{i=1}^{N}\Delta T_{i,j}^{m}-\Delta T_{j}^{o};\label{eq:1}
\end{equation}
and tested  using the spatial t-statistics

\begin{equation}
t_{j}=\frac{\left|\Delta T_{j}\right|}{\sigma_{j}/\sqrt{N}},\label{eq:2}
\end{equation}
where $\sigma_{j}$ is the standard deviation in the cell $j$ among
the $N$ local values $\Delta T_{i,j}^{m}$. A data-model agreement
is rejected (at the two-sided confidence level of 5\%) when $t>t_{0}$,
where $t_{0}$ (the 2.5\% critical value) depends on the degree of
freedom ($N-1$). In our cases, $t_{0}$ is slightly larger than 2.
The proposed statistics optimally emphasizes the regional differences
between observations and a set of simulations.

\begin{table}
{\small{}}%
\begin{tabular}{|l|c|c|c|c|c|c|c|c|c|c|c|}
\hline 
{\small{}CMIP6} & {\small{}ECS} & {\small{}used} & \multicolumn{3}{c|}{{\small{}2011-2021 vs. 1980-1990}} & \multicolumn{3}{c|}{{\small{}2040-2060 vs. 1850-1900}} & \multicolumn{3}{c|}{{\small{}2080-2100 vs. 1850-1900}}\tabularnewline
{\small{}GCM} & {\small{}(°C)} & {\small{}SSP} & {\small{}SSP245} & {\small{}SSP370} & {\small{}SSP585} & {\small{}SSP245} & {\small{}SSP370} & {\small{}SSP585} & {\small{}SSP245} & {\small{}SSP370} & {\small{}SSP585}\tabularnewline
\hline 
\hline 
{\small{}﻿CIESM} & {\small{}5.67} & {\small{}585} & {\small{}0.74} &  & {\small{}0.71} & {\small{}2.72} &  & {\small{}3.07} & {\small{}3.66} &  & {\small{}6.28}\tabularnewline
\hline 
{\small{}CanESM5 p1} & {\small{}5.62} & {\small{}370} & {\small{}1.19} & {\small{}1.20} & {\small{}1.22} & {\small{}3.15} & {\small{}3.51} & {\small{}3.74} & {\small{}4.23} & {\small{}5.97} & {\small{}7.01}\tabularnewline
\hline 
{\small{}CanESM5 p2} & {\small{}5.62} & {\small{}370} & {\small{}1.22} & {\small{}1.22} & {\small{}1.22} & {\small{}3.17} & {\small{}3.56} & {\small{}3.79} & {\small{}4.27} & {\small{}6.04} & {\small{}7.07}\tabularnewline
\hline 
{\small{}CanESM5-CanOE p2} & {\small{}5.62} & {\small{}370} & {\small{}1.16} & {\small{}1.16} & {\small{}1.15} & {\small{}3.2} & {\small{}3.57} & {\small{}3.79} & {\small{}4.22} & {\small{}6.01} & {\small{}7.06}\tabularnewline
\hline 
{\small{}HadGEM3-GC31-LL f3} & {\small{}5.55} & {\small{}585} & {\small{}1.25} &  & {\small{}1.08} & {\small{}2.61} &  & {\small{}3.07} & {\small{}3.81} &  & {\small{}6.12}\tabularnewline
\hline 
{\small{}HadGEM3-GC31-MM-f3} & {\small{}5.42} & 585 &  &  & {\small{}0.86} &  &  & {\small{}3.01} &  &  & {\small{}5.99}\tabularnewline
\hline 
{\small{}UKESM1-0-LL f2} & {\small{}5.34} & {\small{}370} & {\small{}1.11} & {\small{}1.09} & {\small{}1.11} & {\small{}2.76} & {\small{}2.99} & {\small{}3.22} & {\small{}3.97} & {\small{}5.48} & {\small{}6.45}\tabularnewline
\hline 
{\small{}CESM2} & {\small{}5.16} & {\small{}370} & {\small{}0.75} & {\small{}0.76} & {\small{}0.80} & {\small{}2.2} & {\small{}2.28} & {\small{}2.77} & {\small{}3.22} & {\small{}4.17} & {\small{}5.52}\tabularnewline
\hline 
{\small{}CNRM-CM6-1 f2} & {\small{}4.83} & {\small{}370} & {\small{}0.65} & {\small{}0.66} & {\small{}0.68} & {\small{}2.12} & {\small{}2.24} & {\small{}2.54} & {\small{}3.28} & {\small{}4.41} & {\small{}5.65}\tabularnewline
\hline 
{\small{}CNRM-ESM2-1 f2} & {\small{}4.76} & {\small{}370} & {\small{}0.62} & {\small{}0.62} & {\small{}0.65} & {\small{}1.97} & {\small{}2.05} & {\small{}2.32} & {\small{}3.10} & {\small{}4.14} & {\small{}5.01}\tabularnewline
\hline 
{\small{}CESM2-WACCM} & {\small{}4.75} & {\small{}370} & {\small{}0.91} & {\small{}0.88} & {\small{}0.93} & {\small{}2.32} & {\small{}2.40} & {\small{}2.78} & {\small{}3.30} & {\small{}4.25} & {\small{}5.56}\tabularnewline
\hline 
{\small{}ACCESS-CM2} & {\small{}4.72} & {\small{}370} & {\small{}0.82} & {\small{}0.87} & {\small{}0.86} & {\small{}2.35} & {\small{}2.44} & {\small{}2.63} & {\small{}3.41} & {\small{}4.43} & {\small{}5.44}\tabularnewline
\hline 
{\small{}NESM3} & {\small{}4.72} & {\small{}585} & {\small{}0.95} &  & {\small{}1.02} & {\small{}2.13} &  & {\small{}2.75} & {\small{}2.88} &  & {\small{}4.98}\tabularnewline
\hline 
{\small{}IPSL-CM6A-LR} & {\small{}4.56} & {\small{}370} & {\small{}0.76} & {\small{}0.76} & {\small{}0.76} & {\small{}2.55} & {\small{}2.74} & {\small{}2.97} & {\small{}3.58} & {\small{}4.91} & {\small{}5.97}\tabularnewline
\hline 
\hline 
{\small{}Average} & {\small{}5.17} &  & {\small{}0.93} & {\small{}0.92} & {\small{}0.93} & {\small{}2.56} & {\small{}2.78} & {\small{}3.03} & {\small{}3.61} & {\small{}4.98} & {\small{}6.01}\tabularnewline
\hline 
{\small{}Std. Dev.} & {\small{}0.42} &  & {\small{}0.23} & {\small{}0.23} & {\small{}0.20} & {\small{}0.42} & {\small{}0.59} & {\small{}0.47} & {\small{}0.46} & {\small{}0.81} & {\small{}0.71}\tabularnewline
\hline 
\hline 
{\small{}KACE-1-0-G} & {\small{}4.48} & {\small{}370} & {\small{}0.95} & {\small{}0.93} & {\small{}0.96} & {\small{}3.01} & {\small{}3.13} & {\small{}3.34} & {\small{}3.74} & {\small{}4.85} & {\small{}5.67}\tabularnewline
\hline 
{\small{}EC-Earth3-Veg} & {\small{}4.31} & {\small{}370} & {\small{}0.86} & {\small{}0.84} & {\small{}0.87} & {\small{}2.44} & {\small{}2.52} & {\small{}2.80} & {\small{}3.44} & {\small{}4.56} & {\small{}5.42}\tabularnewline
\hline 
{\small{}EC-Earth3} & {\small{}4.3} & {\small{}370} & {\small{}0.76} & {\small{}0.89} & {\small{}0.88} & {\small{}2.34} & {\small{}2.53} & {\small{}2.73} & {\small{}3.39} & {\small{}4.50} & {\small{}5.44}\tabularnewline
\hline 
{\small{}CNRM-CM6-1-HR f2} & {\small{}4.28} & {\small{}370} & {\small{}0.71} & {\small{}0.71} & {\small{}0.73} & {\small{}2.62} & {\small{}2.68} & {\small{}2.96} & {\small{}3.82} & {\small{}4.71} & {\small{}5.76}\tabularnewline
\hline 
{\small{}GFDL-ESM4} & {\small{}3.9} & {\small{}370} & {\small{}0.73} & {\small{}0.67} & {\small{}0.66} & {\small{}1.65} & {\small{}1.75} & {\small{}1.90} & {\small{}2.28} & {\small{}3.24} & {\small{}3.70}\tabularnewline
\hline 
{\small{}ACCESS-ESM1-5} & {\small{}3.87} & {\small{}370} & {\small{}0.83} & {\small{}0.85} & {\small{}0.82} & {\small{}2.09} & {\small{}2.17} & {\small{}2.46} & {\small{}2.99} & {\small{}3.92} & {\small{}4.64}\tabularnewline
\hline 
{\small{}MCM-UA-1-0} & {\small{}3.65} & {\small{}370} & {\small{}0.83} & {\small{}0.79} & {\small{}0.90} & {\small{}2.08} & {\small{}2.20} & {\small{}2.63} & {\small{}2.96} & {\small{}3.89} & {\small{}4.68}\tabularnewline
\hline 
{\small{}CMCC-CM2-SR5} & {\small{}3.52} & {\small{}370} & {\small{}0.61} & {\small{}0.68} & {\small{}0.68} & {\small{}2.45} & {\small{}2.52} & {\small{}2.89} & {\small{}3.62} & {\small{}4.17} & {\small{}5.28}\tabularnewline
\hline 
{\small{}AWI-CM-1-1-MR} & {\small{}3.16} & {\small{}370} & {\small{}0.79} & {\small{}0.76} & {\small{}0.78} & {\small{}2.3} & {\small{}2.45} & {\small{}2.52} & {\small{}2.93} & {\small{}3.98} & {\small{}4.69}\tabularnewline
\hline 
{\small{}MRI-ESM2-0} & {\small{}3.15} & {\small{}370} & {\small{}0.71} & {\small{}0.63} & {\small{}0.80} & {\small{}2.04} & {\small{}2.13} & {\small{}2.41} & {\small{}2.70} & {\small{}3.53} & {\small{}4.31}\tabularnewline
\hline 
{\small{}BCC-CSM2-MR} & {\small{}3.04} & {\small{}370} & {\small{}0.64} & {\small{}0.65} & {\small{}0.66} & {\small{}1.87} & {\small{}2.14} & {\small{}2.32} & {\small{}2.45} & {\small{}3.64} & {\small{}3.95}\tabularnewline
\hline 
\hline 
{\small{}Average} & {\small{}3.79} &  & {\small{}0.76} & {\small{}0.76} & {\small{}0.79} & {\small{}2.26} & {\small{}2.38} & {\small{}2.63} & {\small{}3.12} & {\small{}4.09} & {\small{}4.87}\tabularnewline
\hline 
{\small{}Std. Dev.} & {\small{}0.52} &  & {\small{}0.10} & {\small{}0.10} & {\small{}0.10} & {\small{}0.37} & {\small{}0.36} & {\small{}0.38} & {\small{}0.52} & {\small{}0.52} & {\small{}0.70}\tabularnewline
\hline 
\hline 
{\small{}FGOALS-f3-L} & {\small{}3} & {\small{}370} & {\small{}0.68} & {\small{}0.70} & {\small{}0.69} & {\small{}2.24} & {\small{}2.44} & {\small{}2.58} & {\small{}2.83} & {\small{}3.89} & {\small{}4.62}\tabularnewline
\hline 
{\small{}MPI-ESM1-2-LR} & {\small{}3} & {\small{}370} & {\small{}0.57} & {\small{}0.57} & {\small{}0.55} & {\small{}1.82} & {\small{}2.00} & {\small{}2.10} & {\small{}2.40} & {\small{}3.38} & {\small{}3.98}\tabularnewline
\hline 
{\small{}MPI-ESM1-2-HR} & {\small{}2.98} & {\small{}370} & {\small{}0.57} & {\small{}0.59} & {\small{}0.57} & {\small{}1.8} & {\small{}2.00} & {\small{}2.07} & {\small{}2.44} & {\small{}3.38} & {\small{}3.92}\tabularnewline
\hline 
{\small{}FGOALS-g3} & {\small{}2.88} & {\small{}370} & {\small{}0.61} & {\small{}0.59} & {\small{}0.60} & {\small{}1.82} & {\small{}2.11} & {\small{}2.16} & {\small{}2.24} & {\small{}3.30} & {\small{}3.62}\tabularnewline
\hline 
{\small{}GISS-E2-1-G p1} & {\small{}2.72} & {\small{}370} &  & {\small{}0.70} &  &  & {\small{}2.12} &  &  & {\small{}3.39} & \tabularnewline
\hline 
{\small{}GISS-E2-1-G p3} & {\small{}2.72} & {\small{}370} & {\small{}0.58} & {\small{}0.40} & {\small{}0.45} & {\small{}1.97} & {\small{}2.10} & {\small{}2.33} & {\small{}2.58} & {\small{}3.46} & {\small{}4.10}\tabularnewline
\hline 
{\small{}MIROC-ES2L f2} & {\small{}2.68} & {\small{}370} & {\small{}0.59} & {\small{}0.56} & {\small{}0.56} & {\small{}1.75} & {\small{}1.86} & {\small{}2.15} & {\small{}2.42} & {\small{}3.20} & {\small{}3.97}\tabularnewline
\hline 
{\small{}MIROC6} & {\small{}2.61} & {\small{}370} & {\small{}0.48} & {\small{}0.47} & {\small{}0.51} & {\small{}1.57} & {\small{}1.70} & {\small{}1.93} & {\small{}2.16} & {\small{}2.96} & {\small{}3.75}\tabularnewline
\hline 
{\small{}NorESM2-LM} & {\small{}2.54} & {\small{}370} & {\small{}0.62} & {\small{}0.76} & {\small{}0.71} & {\small{}1.35} & {\small{}1.48} & {\small{}1.76} & {\small{}1.96} & {\small{}2.86} & {\small{}3.69}\tabularnewline
\hline 
{\small{}NorESM2-MM} & {\small{}2.5} & {\small{}370} & {\small{}0.77} & {\small{}0.67} & {\small{}0.71} & {\small{}1.56} & {\small{}1.61} & {\small{}1.86} & {\small{}2.12} & {\small{}3.00} & {\small{}3.77}\tabularnewline
\hline 
{\small{}CAMS-CSM1-0} & {\small{}2.29} & {\small{}370} & {\small{}0.42} & {\small{}0.38} & {\small{}0.41} & {\small{}1.43} & {\small{}1.59} & {\small{}1.64} & {\small{}1.98} & {\small{}2.69} & {\small{}3.11}\tabularnewline
\hline 
{\small{}INM-CM5-0} & {\small{}1.92} & {\small{}370} & {\small{}0.59} & {\small{}0.65} & {\small{}0.60} & {\small{}1.66} & {\small{}1.94} & {\small{}2.15} & {\small{}2.27} & {\small{}3.20} & {\small{}3.59}\tabularnewline
\hline 
{\small{}INM-CM4-8} & {\small{}1.83} & {\small{}370} & {\small{}0.54} & {\small{}0.54} & {\small{}0.59} & {\small{}1.81} & {\small{}1.92} & {\small{}2.17} & {\small{}2.25} & {\small{}3.23} & {\small{}3.74}\tabularnewline
\hline 
\hline 
{\small{}Average} & {\small{}2.59} &  & {\small{}0.58} & {\small{}0.59} & {\small{}0.58} & {\small{}1.73} & {\small{}1.91} & {\small{}2.07} & {\small{}2.31} & {\small{}3.23} & {\small{}3.82}\tabularnewline
\hline 
{\small{}Std. Dev.} & {\small{}0.38} &  & {\small{}0.09} & {\small{}0.12} & {\small{}0.09} & {\small{}0.24} & {\small{}0.26} & {\small{}0.25} & {\small{}0.25} & {\small{}0.30} & {\small{}0.36}\tabularnewline
\hline 
\end{tabular}{\small\par}

\caption{(c1) The adopted CMIP6 GCMs; (c2) their ECS; (c3) SSP simulation analyzed
in Section 3; (c3-c5) 2011-2021 global surface warming (°C, 1980-1990
anomalies); (c6-c8) 2040-2060 and (c9-c11) 2080-2100 global surface
warming (°C, 1850-1900 anomalies).}
\label{tab1}
\end{table}

\begin{figure}[!t]
\begin{centering}
\includegraphics[width=1\textwidth]{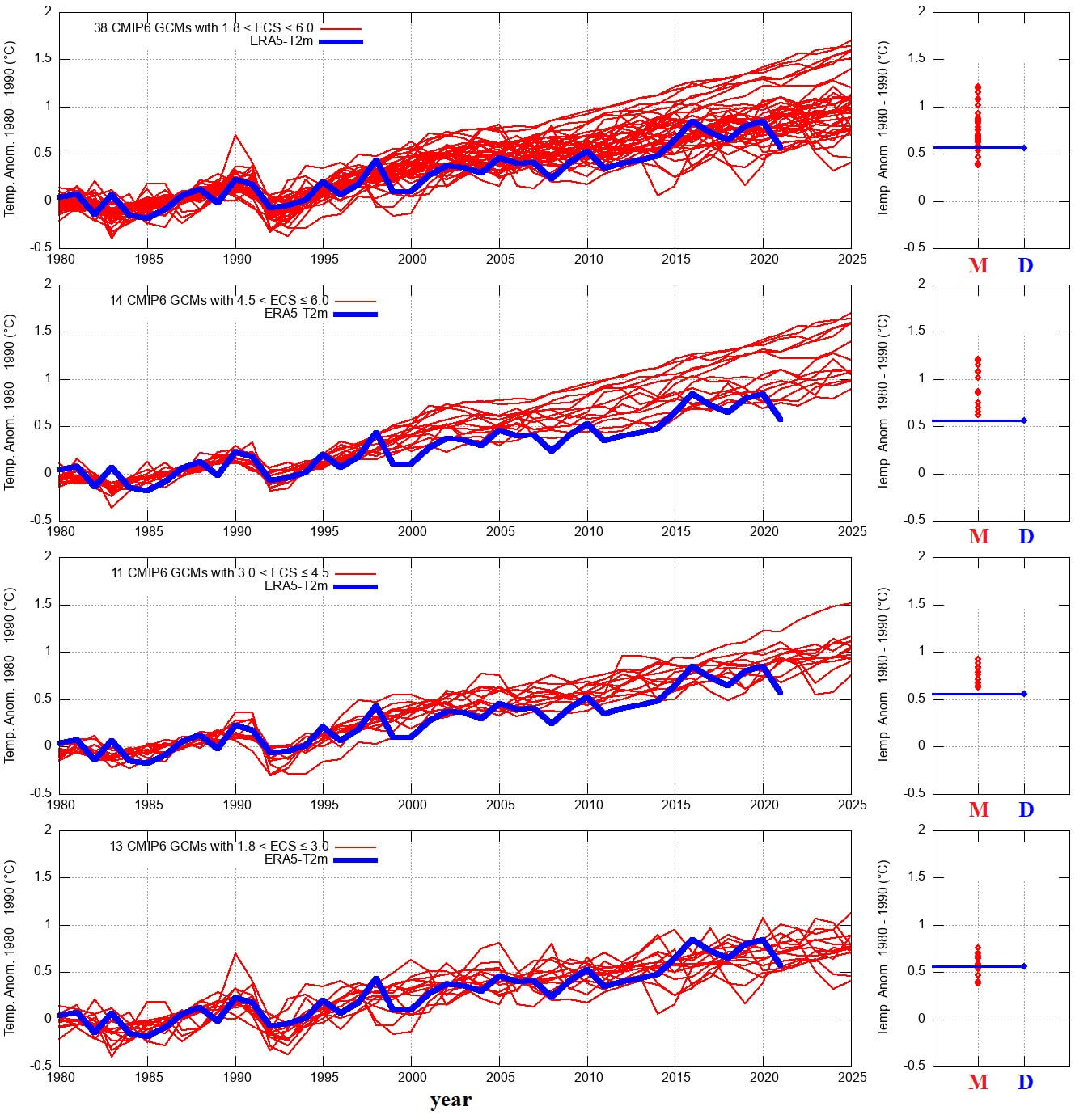}
\par\end{centering}
\centering{}\caption{Left: Global surface temperature simulations (red) against ERA5-T2m
(blue). Right: modeled (M, red) and observed (D, blue, $\Delta T\approx0.56$
°C) mean global surface warming 2011-2021 (1980-1990 anomalies).}
\label{Fig1}
\end{figure}

\section{Results}

This section analyzes 38 CMIP6 simulations referring to the SSP scenario
listed in Table \ref{tab1}-c3, one for each model.

Figure \ref{Fig1} shows the 38 synthetic global surface temperature
records (1980-1990 anomalies, red curves) against ERA5-T2m (blue).
Three panels depict separately the records from 13 low-ECS, 11 medium-ECS,
and 14 high-ECS GCMs. The right panels compare the 2011-2021 mean
temperatures of the models (red dots) against the data (blue dots)
and show that only the low-ECS GCMs are consistent with the observed
warming, while both the medium and high-ECS GCMs definitely exceed
it.

\begin{figure}[!t]
\begin{centering}
\includegraphics[width=1\textwidth]{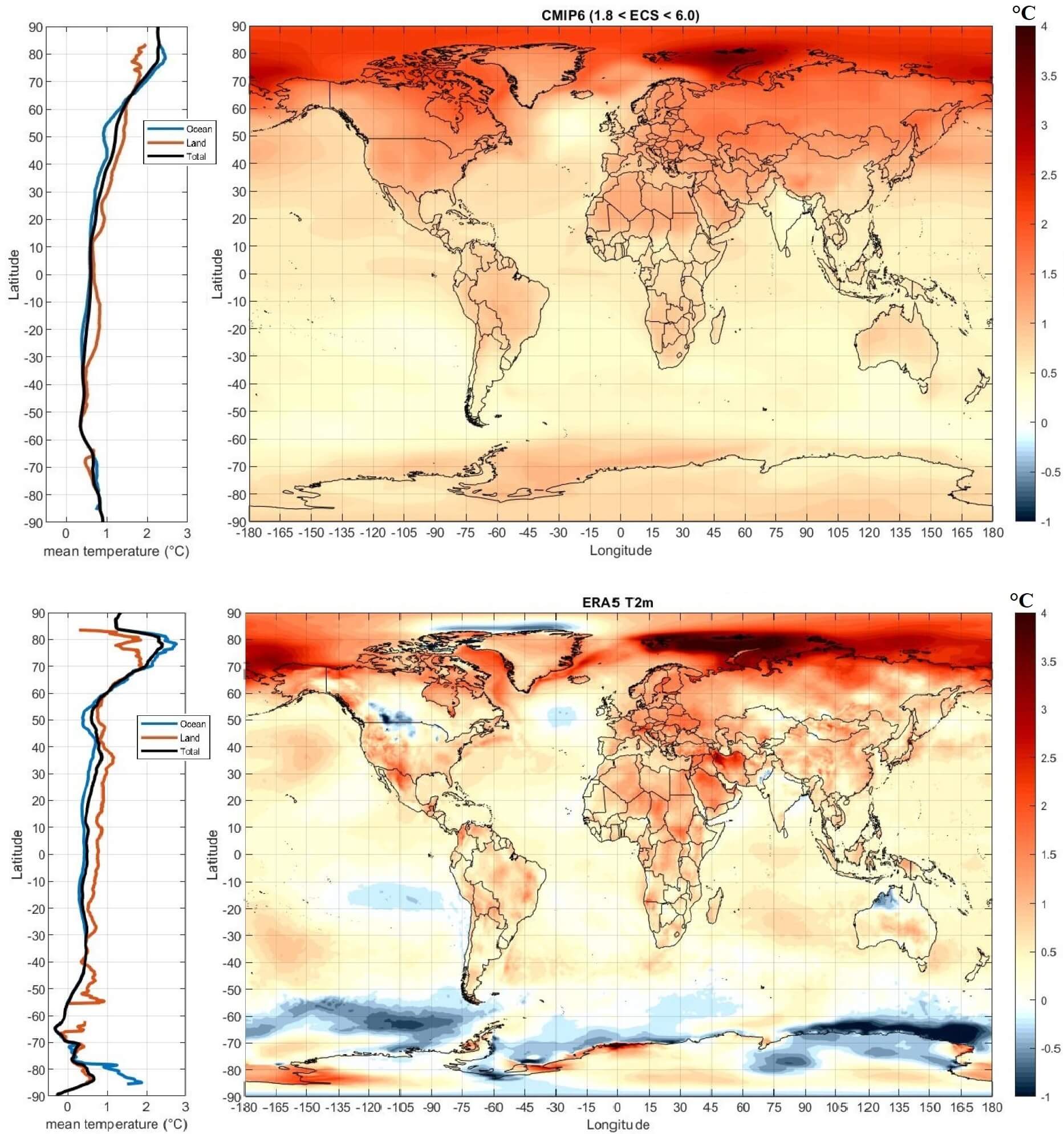}
\par\end{centering}
\centering{}\caption{Warming patterns from 1980-1990 to 2011-2021 predicted by the CMIP6
tas ensemble average record (top) and ERA5-T2m (bottom). Left-sides:
latitudinal temperature profiles for land, ocean and land+ocean regions.}
\label{Fig2}
\end{figure}

Figure \ref{Fig2} shows the world distributions of the 2011-2021
warming (1980-1990 anomalies) of the GCM ensemble mean (top panel)
and the ERA5-T2m record (bottom panel), and the relative latitudinal
temperature profiles (left panels) for the land (brown), ocean (blue)
and land+ocean (black) regions. The simulation predicts a worldwide
diffused warming. The high latitudes (60-90°N) and the continents
generally warmed more than other regions because of albedo changes
related to sea-ice melting and the lower heat capacity of land versus
ocean. ERA5-T2m too shows a diffused warming, however, contrary to
the simulation, it also presents vast cooling regions (blue) around
Antarctica, over the tropical Pacific and North Atlantic oceans, and
over some land regions between the USA and Canada and in North-West
Australia.

\begin{figure}[!t]
\centering{}\includegraphics[width=1\textwidth]{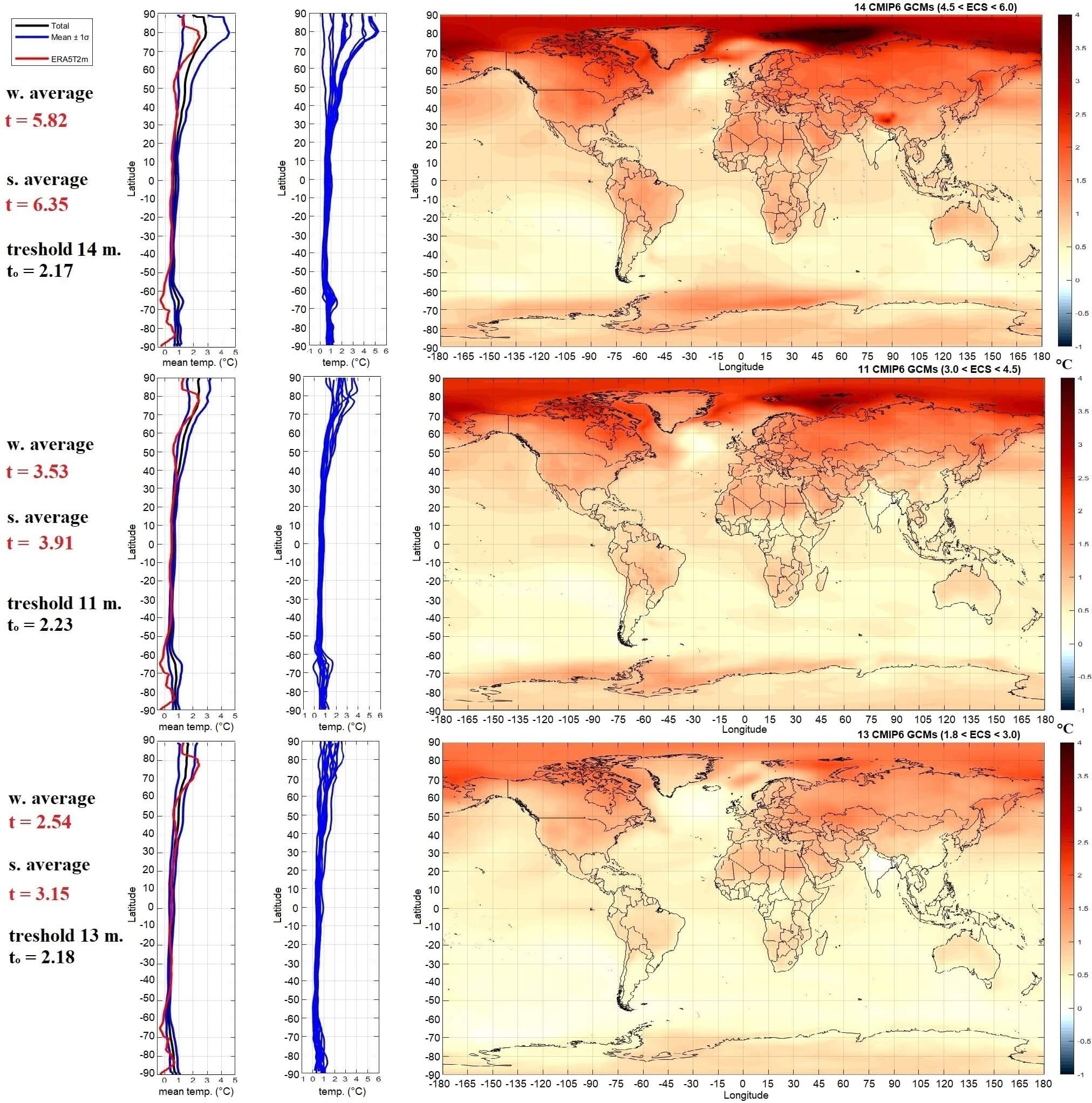}\caption{Spatial warming patterns from 1980-1990 to 2011-2021 produced by high,
medium and low-ECS GCMs. Left-panels: latitudinal temperature profiles
for each model (blue curves); their statistical distribution against
the ERA5-T2m (red) profiles; the correspondent simple and weighted
mean Student's $t$ values and threshold limit $t_{0}$ (Eq. \ref{eq:2}).}
\label{Fig3}
\end{figure}

Figure \ref{Fig3} shows the temperature maps produced by the high,
medium and low-ECS GCMs, the latitudinal profiles of the single GCMs
(blue curves), and their distribution (black, mean curve; blue, $\pm\sigma$
range) versus the ERA5-T2m latitudinal temperature profile  (red).
The figure also shows the correspondent means of the t-test values
 (Eq. \ref{eq:2}) calculated from both the latitudinal profiles using
the mean and a latitudinal-cosine weighted mean, and the correspondent
threshold value $t_{0}$ (black). A statistical model-data agreement
is not found (at the confidence level of 5\%), although the low-ECS
GCMs get the closest to the data ($t=2.54>t_{0}=2.18$).

\begin{figure}[!t]
\centering{}\includegraphics[width=1\textwidth]{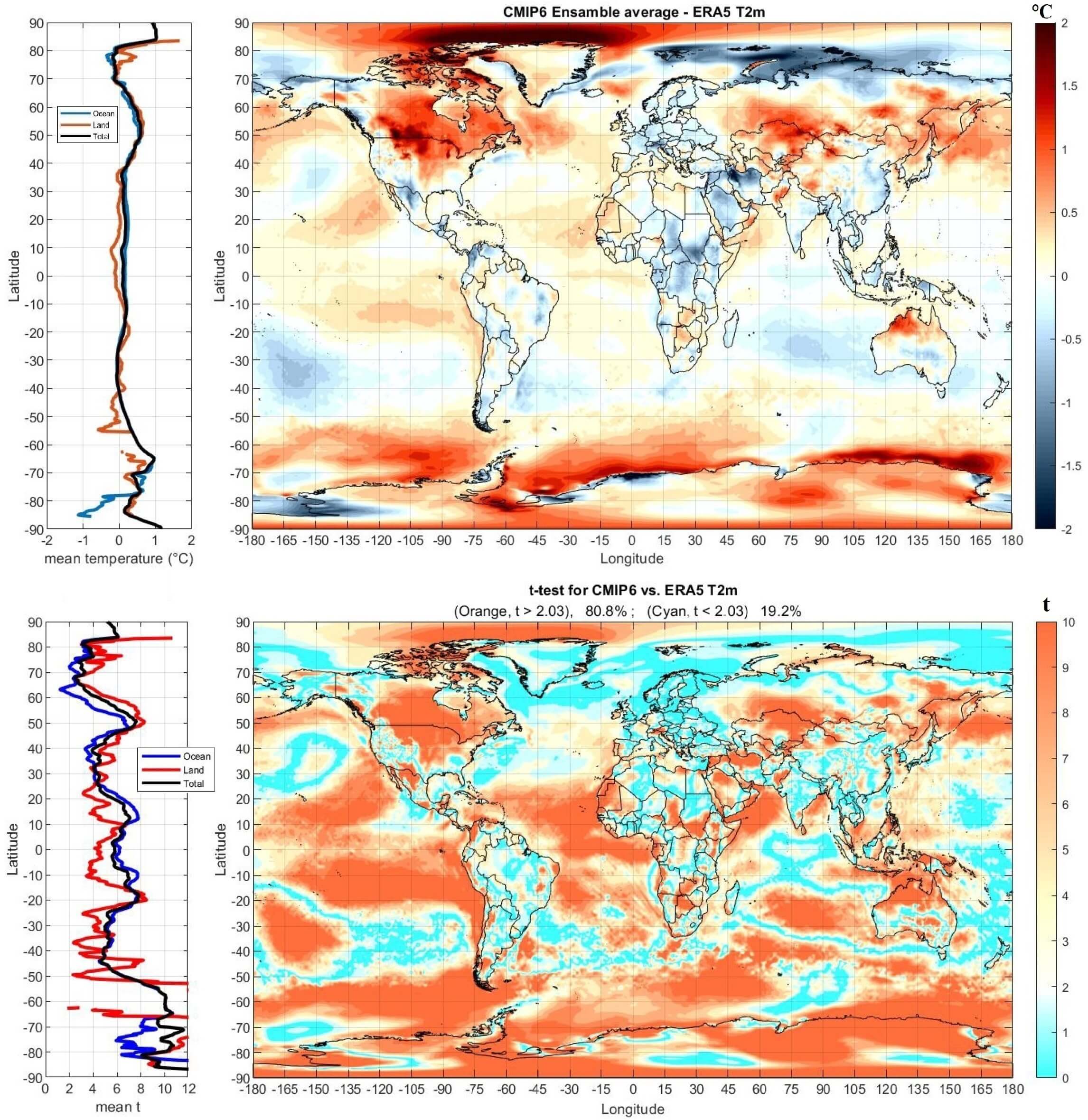}\caption{(Top) Differences between the GCM ensemble average tsa and ERA5-T2m
(Eq. \ref{eq:1}). (Bottom) Spatial t-statistics between the 38 models
and ERA5-T2m (Eq. \ref{eq:2}). Model-data compatibility (at the confidence
level of 5\%) occurs in the cyan color areas ($t<t_{0}=2.03$) while
yellow-red areas indicate statistical incompatibility ($t>t_{0}=2.03$).
Left-panels: relative latitudinal profiles.}
\label{Fig4}
\end{figure}

Figure \ref{Fig4} (bottom) shows the spatial t-value distribution.
The statistics  rejects the simulations over 81\% of the Earth's surface.
Statistical compatibility occurs in the cyan areas ($t<t_{0}$), yellow-red
areas indicate  incompatibility ($t>t_{0}$). Several interesting
patterns are observed. For example, a better  agreement appears in
the Northern Hemisphere and on the continents, although this might
be coincidental because land regions are likely affected by non-climatic
warming biases \citep[cf.:][]{Connolly,Tole,Ouyang2019,Scafetta2021.2}.
The spatial t-statistics also emphasizes a large  divergence over
the oceanic  main current gyres, in particular over the inter-tropical
Pacific and Atlantic, and the Antarctic Circumpolar region. The result
complements \citet{ScafettaMDPI2021} where single-model simulations,
and less robust ensemble statistics were adopted.

\begin{figure}[!t]
\centering{}\includegraphics[width=1\textwidth]{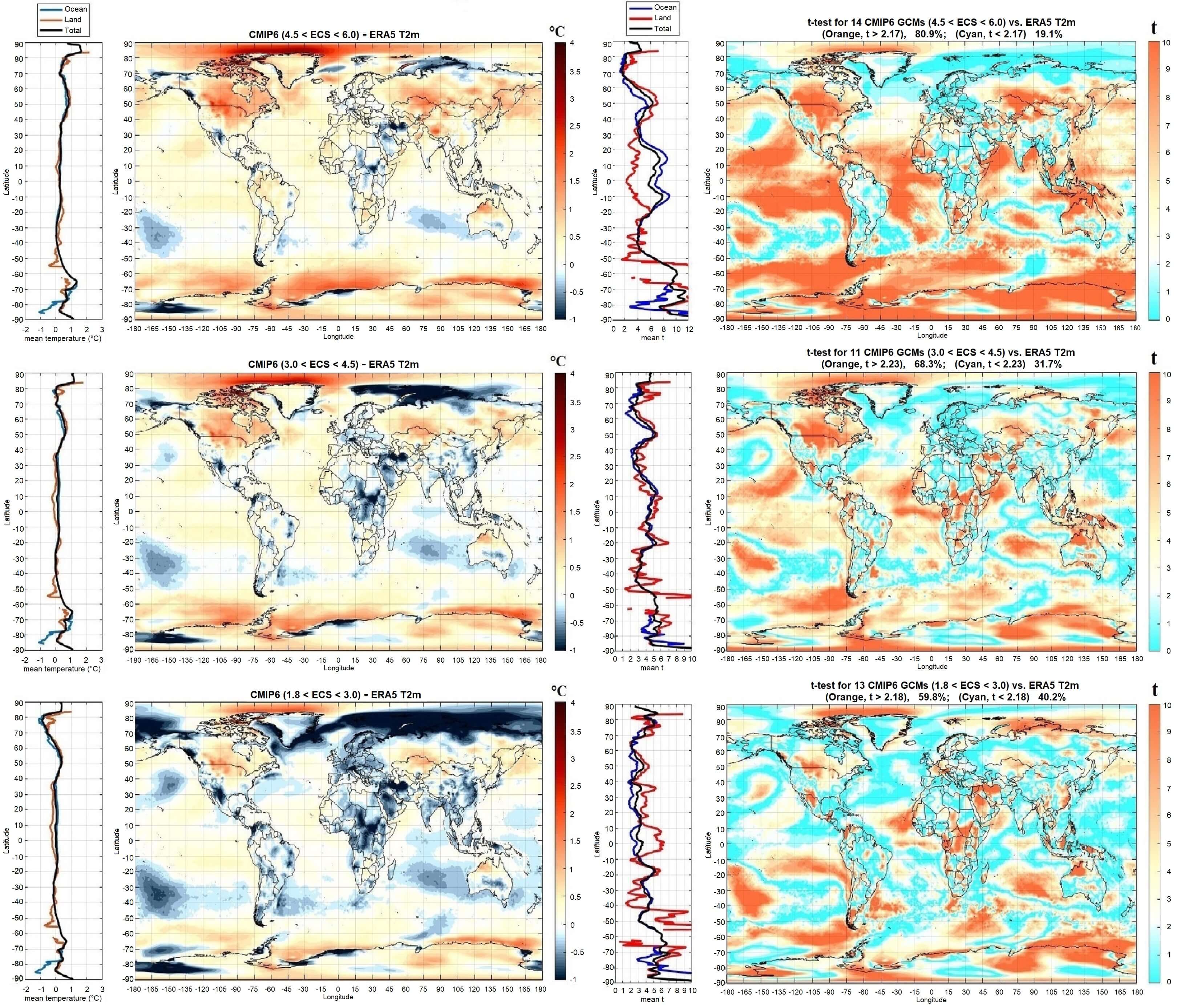}\caption{(Left) Differences between the high, medium and low-ECS GCM ensemble
averages and ERA5-T2m (Eq. \ref{eq:1}). (Right) Spatial t-statistics
between high, medium and low-ECS GCMs and ERA5-T2m (Eq. \ref{eq:2}).
Left-panels: relative latitudinal profiles. Model-data compatibility
(at the significance level of 5\%) occurs in the cyan color areas
($t<t_{0}$) while yellow-red areas indicate statistical incompatibility
($t>t_{0}$).}
\label{Fig5}
\end{figure}

Figure \ref{Fig5} (left panels) shows the differences between the
high, medium and low-ECS GCMs and ERA5-T2m, and their relative latitudinal
temperature profiles. The high-ECS GCMs overestimate the warming over
most of the globe (cf. Figure \ref{Fig1}). The situation slightly
improves with the medium-ECS GCMs. Yet, it remains unsatisfactory
also using the low-ECS GCMs because, for example, the ocean warming
generally appears overestimated while the land one underestimated.
Figure \ref{Fig5} (right panels) shows the spatial t-statistics between
the high, medium and low-ECS GCMs versus ERA5-T2m with their relative
latitudinal temperature profiles. The t-test rejects the agreement
over 60\% (using the low-ECS GCMs) to 81\% (using the high-ECS GCMs)
of the Earth's surface.

\section{Discussion and implication for policy}

It has previously been shown that the model trends of the last decades
are too steep both at the surface and in the troposphere \citep[e.g.:][]{McKitrick,McKitrick(2020),Mitchell,Scafetta2013,Scafetta2021,ScafettaMDPI2021,Tokarska,Wang}.
This analysis focuses on the 1980-2021 period at the surface level
and shows that, while no model group succeeds in reproducing the observed
surface warming patterns, the high ECS models  systematically do worse.
Figure \ref{Fig1} compares the surface temperature changes for stratified
 ECS values and shows that medium and high ECS GCMs exceed the observed
warming. Figures 2-5 suggest where, on the globe, the problems may
be greatest. However, the model-data agreement between  CMIP6 GCMs
and ERA5-T2m is in general very poor both locally and globally,  suggesting
that the global trend alignment with the low-ECS GCMs (Figure \ref{Fig1})
may be coincidental because it is not supported by the regional analysis
(Figure \ref{Fig5}).

Accurately reproducing regional temperature differences over the past
40+ years is beyond the capability of climate model simulations, and
 even fails for major ocean basins and continents. The result suggests
the existence of major issues with all models and/or with the ERA5-T2m
record, which may also be affected by uncorrected local non-climatic
biases \citep[e.g. over land:][]{Connolly,Tole,Ouyang2019,Scafetta2021.2}.

\begin{figure}[!t]
\begin{centering}
\includegraphics[width=1\textwidth]{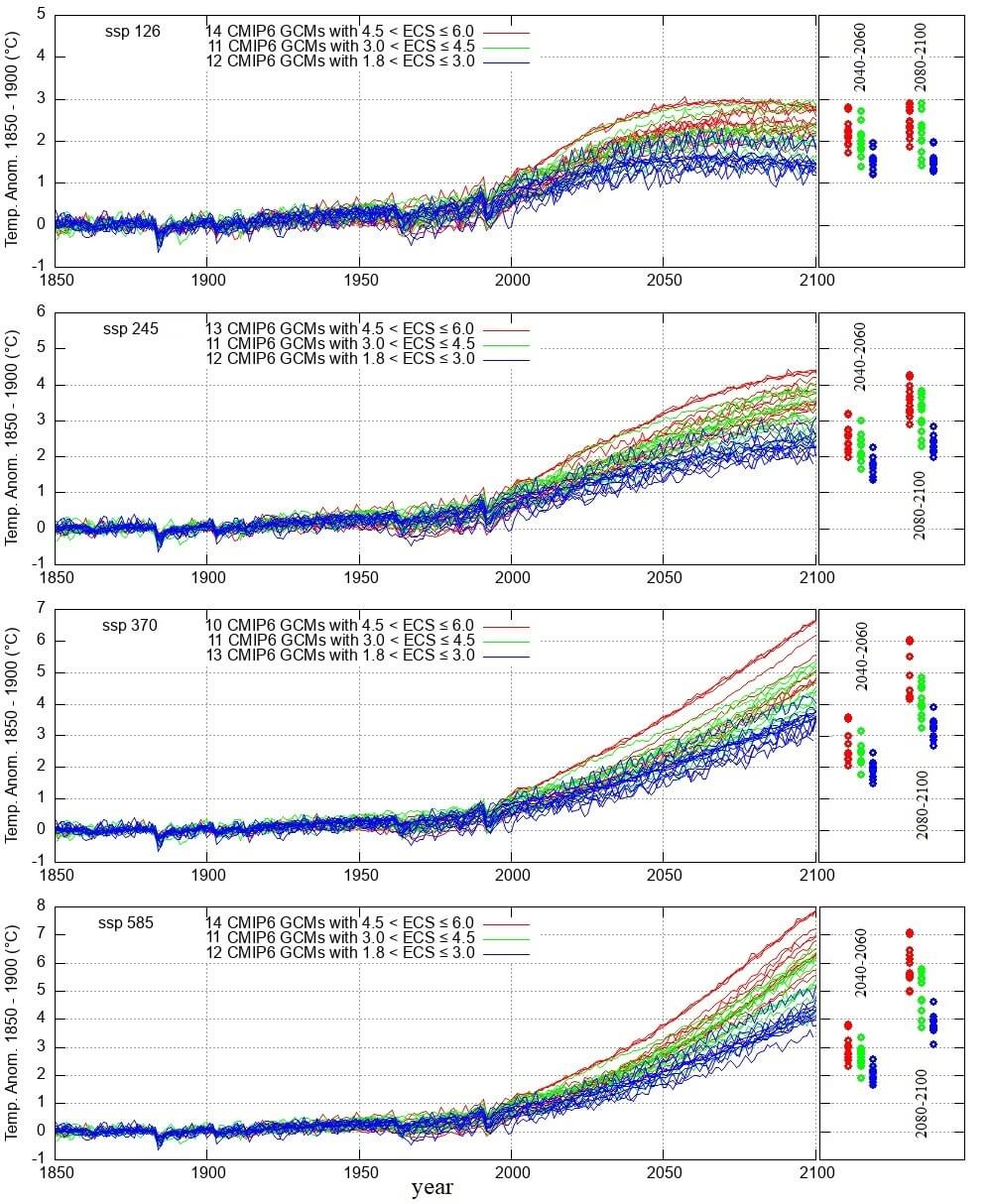}
\par\end{centering}
\centering{}\caption{All available CMIP6 GCM simulations for different SSP: low-ECS (blue),
medium-ECS (green) and high-ECS (red) GCMs. Right-panels: model mean
warming levels in 2040-2060 and 2080-2100. See Table \ref{tab1}.}
\label{Fig6}
\end{figure}

The evidence here presented  indicates that the CMIP6 GCMs do not
reproduce well both the global and regional (i.e., country-size) responses
to enhanced greenhouse gases over the past 40+ years, which calls
into question model-based attribution of climate responses to anthropogenic
forcing.

The fact that low-ECS models perform relatively better while representing
post-1980 surface temperature responses has important social and political
implications. Policy plans are  based on projections like those in
Figure \ref{Fig6}, which compares all available CMIP6 GCM simulations
for each available SSP relative to the pre-industrial era (1850-1900)
up to 2100. The curves are  differently colored according to the GCM
ECS value: low-ECS (blue), medium-ECS (green) and high-ECS (red).
The right panels show the correspondent warming levels in 2040-2060
and 2080-2100. The GCM predicted warming increases with ECS.

The prospect of experiencing a high level of warming associated with
high ECS models has led to costly international efforts to reduce
net greenhouse gas emissions to zero on a rapid timetable. However,
the low-ECS GCMs predict average warming by 2040-2060 close to or
lower than 2 °C even for the SSP585 scenario (Table \ref{tab1}),
which is considered highly unlikely \citep{IPCC2021}. The fact that
the high and medium-ECS GCMs do not appear to be consistent with the
observations over the past 40+ years imply that their projections
should not be used as a basis for policy. The low-ECS GCMs are the
closest to the data but they are unalarming because they predict moderate
warming ($\Delta T_{preindustrial\rightarrow2050}\lesssim2\:{^\circ}C$).
Thus, inexpensive adaptation policies should be preferred because
they should be sufficient to address most hazards related to future
climate changes.

\section{Conclusions}

The large ECS range (between 1.8 and 5.7 °C) of the CMIP6 GCMs indicates
that these models are intrinsically very different from each other.
To constrain it, we grouped the models into three different classes
(low-ECS, 1.80-3.00 °C; medium-ECS, 3.01-4.50 °C; and high-ECS, 4.51-6.00
°C) and tested their predicted temperature changes from the 1980-1990
to 2011-2021 periods against ERA5-T2m. The proposed methodology compares
single GCM runs and specific ensemble averages versus the observations
using both the temperature changes  (Eq. \ref{eq:1}) and spatial
t-statistics (Eq. \ref{eq:2}) maps. This technique optimally highlights
both the global and local discrepancies between the observations and
GCM ensemble predictions. Figures \ref{Fig4} and \ref{Fig5} show
applications of the proposed methodology and provide much information
regarding the performance of the CMIP6 GCMs in simulating the data.

High and medium-ECS GCMs overestimate the observed warming. The low-ECS
GCMs appear to agree with the observations better on average (Figure
\ref{Fig1}), but they  still poorly perform when synoptic temperature
patterns are analyzed (Figure \ref{Fig5}). In general, spatial t-statistics
demonstrated that over more than 60\% of the world surface the CMIP6
GCM predictions are incompatible with the temperature records (at
the significance level of 5\%). Various Northern-Southern hemispheric
and land-ocean asymmetries, and important dynamical patterns -- such
as those related to the main oceanic currents of the Pacific, the
Atlantic and around the Antarctic Circumpolar region -- are observed.
The results suggest poor modeling of heat transfer, ocean and atmospheric
circulation, and Arctic and Antarctic sea-ice processes. Furthermore,
overland, ERA5-T2m is likely affected by numerous non-climatic biases
that could have stressed local warming trends \citep{Tole,Ouyang2019,Scafetta2021.2}.

The result has important implications for policy as well because the
medium and high-ECS GCMs are not sufficiently reliable at the largest
scale (global average) and all GCMs fail to provide confident forecasts
of regional responses to enhanced GHGs so that also the impacts of
policy options on the regional climate are highly uncertain. The models
that best match the post-1980 observations imply that by 2050 the
global surface warming should remain moderate ($\Delta T\lesssim2\:{^\circ}C$)
compared to preindustrial temperatures even under an extremely high
emissions growth scenario with no mitigation effort.

\subsection*{Conflict of Interest}

The author declares no conflict of interest.

\subsection*{Data Availability Statement}

Data from KNMI Climate Explorer:
\begin{itemize}
\item ERA5-T2m, \href{http://climexp.knmi.nl/selectfield_rea.cgi?id=someone@somewhere}{http://climexp.knmi.nl/selectfield\_rea.cgi?id=someone@somewhere};
\item CMIP6 GCMs, \href{http://climexp.knmi.nl/selectfield_cmip6.cgi?id=someone@somewhere}{http://climexp.knmi.nl/selectfield\_cmip6.cgi?id=someone@somewhere}.
\end{itemize}

\end{document}